\documentclass{ieeeaccess}
\usepackage{cite}
\usepackage{amsmath,amssymb,amsfonts}
\usepackage{algorithmic}
\usepackage{graphicx}
\usepackage{textcomp}

\usepackage{bm}
\makeatletter
\AtBeginDocument{\DeclareMathVersion{bold}
\SetSymbolFont{operators}{bold}{T1}{times}{b}{n}
\SetSymbolFont{NewLetters}{bold}{T1}{times}{b}{it}
\SetMathAlphabet{\mathrm}{bold}{T1}{times}{b}{n}
\SetMathAlphabet{\mathit}{bold}{T1}{times}{b}{it}
\SetMathAlphabet{\mathbf}{bold}{T1}{times}{b}{n}
\SetMathAlphabet{\mathtt}{bold}{OT1}{pcr}{b}{n}
\SetSymbolFont{symbols}{bold}{OMS}{cmsy}{b}{n}
\renewcommand\boldmath{\@nomath\boldmath\mathversion{bold}}}
\makeatother

\def\BibTeX{{\rm B\kern-.05em{\sc i\kern-.025em b}\kern-.08em
    T\kern-.1667em\lower.7ex\hbox{E}\kern-.125emX}}

\usepackage{algorithm}
\usepackage{multirow}

\begin{document}
\doi{10.1109/ACCESS.2024.0429000}

\title{Improving search efficiency via adaptive acquisition function selection in discrete black-box optimization}
\author{\uppercase{Reo Shikanai}\authorrefmark{1, 2},
\uppercase{Masayuki Ohzeki}\authorrefmark{1,2,3,4}}

\address[1]{Graduate School of Information Sciences, Tohoku University, Sendai, Japan}
\address[2]{Sigma-i Co., Ltd., Tokyo, Japan}
\address[3]{Department of Physics, Institute of Science Tokyo, Tokyo, Japan}
\address[4]{Research and Education Institute for Semiconductors and Informatics, Kumamoto University, Kumamoto, Japan}

\markboth
{Author \headeretal: Preparation of Papers for IEEE TRANSACTIONS and JOURNALS}
{Author \headeretal: Preparation of Papers for IEEE TRANSACTIONS and JOURNALS}

\corresp{Corresponding author: Reo Shikanai (e-mail: reo.shikanai.s1@dc.tohoku.ac.jp).}

\begin{abstract}
In discrete-variable black-box optimization, the number of candidate solutions grows combinatorially, while each evaluation is often expensive. Therefore, it is important to identify promising solutions efficiently within a limited number of trials. Bayesian Optimization of Combinatorial Structures (BOCS), an existing parametric method, works effectively when only a small amount of data is available. However, as the number of observations increases, BOCS tends to repeatedly propose points that have already been evaluated, which leads to search stagnation. A random-point addition strategy has been proposed to address this issue when an evaluated point is proposed, but it cannot sufficiently exploit information from promising data obtained so far. In this study, we propose a hybrid method that uses BOCS as the main search framework and generates alternative unevaluated points using a Gaussian process only when search stagnation is detected. In the Gaussian-process-based component, multiple Lower Confidence Bound (LCB) acquisition functions are adaptively selected to dynamically control the balance between exploitation and exploration. Numerical experiments using fully connected Quadratic Unconstrained Binary Optimization (QUBO) and Higher-order Unconstrained Binary Optimization (HUBO) as black-box functions show that the proposed method finds solutions with better objective values than the conventional random-point addition method in both settings. Additional analyses show that its effectiveness comes from selecting points that promote search progress within Hamming-distance neighborhoods, rather than simply adding low-energy points near promising solutions. Experiments with sparse surrogate models for quantum annealer applications further suggest the importance of retaining near-fully connected representational capacity.
\end{abstract}

\begin{keywords}
Black-box optimization, Combinatorial optimization problem, Gaussian process.
\end{keywords}

\titlepgskip=-21pt

\maketitle

\section{Introduction}

Black-box optimization is a framework for searching for solutions with low or high objective values within a limited number of evaluations when the explicit mathematical form or internal structure of the objective function is unavailable. This study focuses on problems in which the input has a discrete combinatorial structure. In such problems, the number of candidates grows combinatorially, and the evaluation of each candidate is often expensive. Therefore, it is important to reach promising solutions with a small number of trials. Such problems arise in application areas such as chemical materials discovery, metamaterial design, and drug discovery \cite{doi2023chemical,kitai2020designing,tucs2023quantum}. In black-box optimization, it is important to balance exploitation, which investigates known promising regions, and exploration, which investigates uncertain unknown regions. This balance is usually controlled by the design of a surrogate model and an acquisition function that suggests the next search point \cite{jones1998efficient,shahriari2016taking}. However, in discrete-variable spaces, the number of candidates increases exponentially, and thus the representational capacity of the surrogate model and the search strategy of the acquisition function strongly affect the optimization performance.

In discrete-variable black-box optimization, methods based on parametric surrogate models and those based on nonparametric surrogate models are widely used, and each has different advantages and limitations. Bayesian Optimization of Combinatorial Structures (BOCS) is a representative parametric method for discrete black-box optimization \cite{baptista2018bocs}. BOCS constructs a surrogate model that approximates the unknown objective function from past observations and determines the next search point based on the approximated function. Because the surrogate model is represented as a quadratic binary function, BOCS can learn relatively stably even with a small amount of data. On the other hand, its representational capacity is limited. As the number of observations increases, learning may stagnate, and the acquisition function tends to repeatedly propose points that have already been evaluated \cite{morita2023random}. In this study, we refer to this phenomenon as learning stagnation. Nonparametric methods represented by Gaussian processes have high representational capacity and can flexibly approximate the objective function when sufficient observations are available \cite{rasmussen2006gpml}. However, such flexibility does not always lead to better search performance. In fact, previous experiments have shown that Gaussian processes can perform worse than BOCS in terms of search performance \cite{baptista2018bocs}. Nevertheless, it remains unclear whether Gaussian processes are still inferior to BOCS after promising data have been accumulated.

To address learning stagnation in BOCS, a strategy has been proposed in which a random point is added as a postprocessing step when an already evaluated point is proposed \cite{morita2023random}. This strategy has been shown to improve learning stagnation in several problem settings. Although it can explore unknown regions with almost no increase in computational cost, it cannot sufficiently reflect information from the promising regions obtained so far when generating the next point. As a result, the search may move away from the neighborhood of the optimum. Therefore, when learning stagnation occurs, a mechanism is needed that provides more useful additional points by exploiting the accumulated data, rather than simply adding random points.

In this study, we propose a method that uses Gaussian processes to provide promising search points when learning stagnation occurs in BOCS. Specifically, we introduce a hybrid framework in which BOCS efficiently advances the search in the early stage with limited data, and Gaussian processes complement the search only when learning stagnation is detected after observations have accumulated. As described above, Gaussian processes may be effective when sufficient data have been accumulated, which motivates this approach. However, when an acquisition function based on a single Gaussian process is used, parameter tuning is important for controlling the balance between exploitation and exploration. This tuning becomes a practical issue in real-world problems where the number of evaluations is limited. Therefore, we introduce GP-Hedge, which adaptively selects from multiple acquisition functions according to their past performance \cite{hoffman2011portfolio}. Compared with the use of a single acquisition function, GP-Hedge can reduce the burden of parameter tuning, and experiments have shown that it can achieve better search performance than a single Gaussian process. In addition, the computations of multiple acquisition functions can be parallelized independently, which helps suppress the increase in computational time.

This study also investigates performance improvement from the viewpoint of the optimization method used for the acquisition function in BOCS. Because BOCS represents the surrogate model as a quadratic binary function, the acquisition function can be formulated as a Quadratic Unconstrained Binary Optimization (QUBO) problem. For QUBO optimization, quantum annealing has attracted attention in addition to classical heuristics such as simulated annealing. Quantum annealing was proposed as an optimization method that uses quantum fluctuations induced by a transverse magnetic field instead of thermal fluctuations \cite{kirkpatrick1983optimization,kadowaki1998qa}, and hardware implementations were later reported \cite{johnson2011qa}. In recent years, studies have used quantum annealers from D-Wave Quantum Inc. as solvers for the acquisition function in BOCS, and their effectiveness has been investigated in benchmark problems \cite{koshikawa2021dwavebbo}, materials discovery \cite{doi2023chemical}, and data cleansing \cite{otsuka2025filtering}. However, these studies may not fully exploit the performance of quantum annealers.

When a QUBO problem is solved on a quantum annealer, the graph structure of the QUBO must be embedded into a sparse hardware graph \cite{choi2008minor}. In particular, when the target QUBO is dense, computational performance can deteriorate substantially. On the other hand, experiments have shown that approximate solutions can be obtained faster than with classical methods under an ideal setting in which the QUBO graph matches the hardware graph \cite{tasseff2024emerging}. Therefore, to fully exploit the performance of quantum annealers within the BOCS framework, it is important to make the surrogate model more sparsely connected and reduce the effect of embedding. However, such an operation in BOCS leads to a decrease in the representational capacity of the surrogate model and involves a trade-off in which learning performance may degrade. Therefore, this study also investigates how much the proposed method contributes to improving search performance when the surrogate model is sparsified. This analysis provides guidance for the future use of quantum annealers.

The remainder of this paper is organized as follows. The Method section describes learning stagnation in BOCS and the proposed BOCS with GP-Hedge method. The Experimental setup section presents the comparison methods, benchmark problems, evaluation metrics, and sparse surrogate model settings. The Results section reports performance comparisons on fully connected QUBO and fully connected Higher-order Unconstrained Binary Optimization (HUBO) problems and analyzes the behavior during learning stagnation and the effect of sparse surrogate models. Finally, the Discussion and conclusion section discusses the additional analyses and concludes the paper.

\section{Method}

\subsection{Limitations of BOCS}

We consider a discrete-variable black-box optimization problem whose input is a $d$-dimensional binary vector $\bm{x}=(x_1,x_2,\ldots,x_d)\in\{0,1\}^d$. The objective is to find an input $\bm{x}^\ast \in \arg\min_{\bm{x}\in\{0,1\}^d} f(\bm{x})$ that minimizes an unknown objective function $f$ under a limited number of evaluations. The explicit functional form and internal structure of $f(\bm{x})$ are unknown, and the function value at each point can be obtained only by querying the black box. We denote the dataset observed up to iteration $t$ by $\mathcal{D}_t = \{(\bm{x}^{(i)}, y^{(i)})\}_{i=1}^{t}$, where $y^{(i)} = f(\bm{x}^{(i)})$.

BOCS approximates the true objective function by using a surrogate model consisting of linear terms and second-order interaction terms of binary variables. This model is quadratic with respect to $\bm{x}$, but it can be treated as a linear model with respect to the regression coefficients. The surrogate model at iteration $t$ is written as
\begin{equation}
\hat{f}_t(\bm{x}|\bm{\alpha}) =
\alpha_0 +
\sum_{i=1}^{d}\alpha_i x_i +
\sum_{i<j}\alpha_{ij}x_i x_j
=\phi(\bm{x})^\top \bm{\alpha}.
\label{eq:bocs-surrogate}
\end{equation}
Here,
\begin{equation}
\phi(\bm{x})=
\left[
1,\ x_1,\ldots,x_d,\ \{x_i x_j\}_{i<j}
\right]
\end{equation}
is a feature vector consisting of the intercept, linear terms, and second-order interaction terms. The vector $\bm{\alpha}=(\alpha_0,\{\alpha_i\}_{i=1}^{d},\{\alpha_{ij}\}_{i<j}) \in \mathbb{R}^{p}$, where $p=1+d+\binom{d}{2}$, denotes the regression coefficients.

BOCS imposes a horseshoe prior on the regression coefficients $\bm{\alpha}$. This prior is expected to shrink unnecessary coefficients toward $0$ while retaining important coefficients. At iteration $t$, BOCS samples coefficients $\bm{\alpha}^{(t)}$ from the posterior distribution based on the observed data $\mathcal{D}_t$ by Gibbs sampling, and then uses the function $\hat{f}_t(\bm{x}|\bm{\alpha}^{(t)})$ defined by the sampled coefficients as an acquisition function.

The next search point is determined by minimizing this acquisition function over the binary search space. That is,
\begin{equation}
\bm{x}^{\mathrm{BOCS}}_{t+1}
\in
\arg\min_{\bm{x}\in\{0,1\}^d}
\hat{f}_t(\bm{x}|\bm{\alpha}^{(t)}).
\label{eq:bocs-candidate}
\end{equation}
By restricting the model structure to the quadratic form in (1), BOCS can make relatively stable predictions for unobserved points even when only a small amount of data is available. However, as the number of observations increases, BOCS has been reported to select already evaluated points again as search points \cite{morita2023random}. In this study, we refer to this phenomenon as learning stagnation.

A previous study proposed a strategy for avoiding learning stagnation by adding a random point when BOCS proposes an already evaluated point \cite{morita2023random}. However, this strategy does not sufficiently exploit information from the promising regions obtained so far when generating the next point. Therefore, this study develops a method that proposes unevaluated and promising search points by exploiting the observed data when learning stagnation occurs.

\subsection{BOCS with GP-Hedge}

This study introduces a hybrid method that combines BOCS with Gaussian processes to generate more promising search points than random-point addition during learning stagnation. The basic idea is to use BOCS in the early stage, where the amount of data is limited, and to generate alternative unevaluated points by using a Gaussian process only when the acquisition function of BOCS proposes an already evaluated point.

We first define the predictive mean and predictive variance of the Gaussian process based on the observed dataset $\mathcal{D}_t$ as
\begin{align}
\mu_t(\bm{x}) &=
\bm{k}_t(\bm{x})^\top K_t^{-1}\bm{y}_t,
\label{eq:gp-mean} \\
\sigma_t^2(\bm{x}) &=
k(\bm{x}, \bm{x})
-
\bm{k}_t(\bm{x})^\top K_t^{-1}\bm{k}_t(\bm{x}).
\label{eq:gp-var}
\end{align}
Here, $K_t$ is the kernel matrix between observed points, and $\bm{k}_t(\bm{x})$ is the vector of kernel values between the observed points and $\bm{x}$. To use a Gaussian process in a discrete-variable space, a kernel function that properly represents the similarity between binary vectors is required. In this study, we use the Hamming kernel, which reflects the number of component-wise mismatches between two binary vectors:
\begin{equation}
k(\bm{x}, \bm{x}')
=
\exp\left(
-\gamma d_{\mathrm{H}}(\bm{x}, \bm{x}')
\right).
\label{eq:hamming-kernel}
\end{equation}
Here, $d_{\mathrm{H}}(\bm{x}, \bm{x}')$ is the Hamming distance between $\bm{x}$ and $\bm{x}'$, and $\gamma > 0$ is a hyperparameter that controls the decay rate with respect to the distance.

The proposed method uses GP-Hedge, which adaptively selects an acquisition function from multiple candidates according to their past performance, instead of fixing a single acquisition function \cite{hoffman2011portfolio}. Although GP-Hedge can handle arbitrary acquisition functions as candidates, this study constructs the candidate set only from Lower Confidence Bound (LCB) acquisition functions to examine the effectiveness of Gaussian-process-based alternative point generation \cite{srinivas2010gpucb}. Specifically, for each acquisition function $m=1,\ldots,M$, we set a different hyperparameter $\kappa_m > 0$ and use the following LCB acquisition function:
\begin{equation}
a_m(\bm{x}) = \mu_{t-1}(\bm{x}) - \kappa_m \sigma_{t-1}(\bm{x}).
\label{eq:lcb}
\end{equation}
A larger $\kappa_m$ increases the contribution of the predictive standard deviation $\sigma_{t-1}(\bm{x})$, and thus tends to prioritize points with high uncertainty. In contrast, a smaller $\kappa_m$ gives relatively more weight to the predictive mean $\mu_{t-1}(\bm{x})$, and thus tends to emphasize known promising regions. Therefore, by varying $\kappa_m$, we can construct a candidate set of acquisition functions with different behaviors, ranging from exploitation-oriented to exploration-oriented search.

Hereafter, following the analogy with multi-armed bandit problems, each acquisition function in GP-Hedge is called an arm. GP-Hedge selects the next acquisition function from these arms based on their past performance. Let $r_t^{(m)}$ be the reward obtained by arm $m$ at iteration $t$, and let $g_t^{(m)} = \sum_{s=1}^{t} r_s^{(m)}$ be the cumulative gain, defined as the sum of past rewards. The initial cumulative gain is set to $g_0^{(m)}=0$ for all arms. In this study, for the candidate point $\tilde{\bm{x}}_t^{(m)}$ recommended by arm $m$ during stagnation, we define the reward as $r_t^{(m)} = -\mu_t\!\left(\tilde{\bm{x}}_t^{(m)}\right)$ by using the GP predictive mean after adding the new observation. Thus, an arm that recommends a candidate with a smaller predictive mean after the GP update tends to obtain a larger cumulative gain. Let $\mathcal{M}_t$ be the set of arms that propose unevaluated points. Then, an arm $m \in \mathcal{M}_t$ is selected with probability
\begin{equation}
p_t(m)=\frac{\exp\!\left(\eta g_{t-1}^{(m)}\right)}{\sum_{\ell \in \mathcal{M}_t}\exp\!\left(\eta g_{t-1}^{(\ell)}\right)}.
\label{eq:hedge-prob}
\end{equation}
Here, $\eta > 0$ is a Hedge parameter that controls how strongly differences in cumulative gains are reflected in the selection probability. As a result, arms that have recommended more promising candidates in the past are more likely to be selected. The restriction to arms that propose unevaluated points is a procedure introduced in this study. Algorithm~\ref{alg:gph} shows the GP-Hedge candidate selection procedure used in one iteration.

Next, Algorithm~\ref{alg:hybrid} shows the overall proposed method that combines BOCS and GP-Hedge. At each iteration, the method first generates a candidate point by using the acquisition function of BOCS. If the candidate is unevaluated, the method evaluates it in the same way as standard BOCS. If BOCS proposes an already evaluated point, the method regards this as learning stagnation and activates GP-Hedge through Algorithm~\ref{alg:gph} to generate an alternative unevaluated point. This structure allows the method to exploit the efficiency of BOCS in the early stage and switch to Gaussian-process-based search only during stagnation. Unlike simple random-point addition, the proposed method can generate alternative points by taking into account the observations obtained by BOCS.

\begin{algorithm}[!htbp]
\caption{GP-Hedge candidate selection used in the stagnation phase}
\label{alg:gph}
\begin{algorithmic}[1]
\STATE \textbf{Input:} observed dataset $\mathcal{D}_{t-1}$, acquisition functions $\{a_m\}_{m=1}^{M}$, Hedge parameter $\eta > 0$, cumulative gains $\{g_{t-1}^{(m)}\}_{m=1}^{M}$
\FOR{$m = 1,\ldots,M$}
    \STATE Nominate $\tilde{\bm{x}}_t^{(m)} \in \arg\min_{\bm{x}\in\{0,1\}^d} a_m(\bm{x})$
\ENDFOR
\STATE Let $\mathcal{M}_t = \{\, m \in \{1,\ldots,M\} \mid \tilde{\bm{x}}_t^{(m)} \notin \{\bm{x}^{(i)}\}_{i=1}^{t-1} \,\}$
\IF{$\mathcal{M}_t \neq \varnothing$}
    \STATE Select arm $j_t \in \mathcal{M}_t$ with probability
    \[
    p_t(m)=\frac{\exp(\eta g_{t-1}^{(m)})}{\sum_{\ell\in\mathcal{M}_t}\exp(\eta g_{t-1}^{(\ell)})}, \quad m\in\mathcal{M}_t
    \]
    \STATE Set $\bm{x}_t = \tilde{\bm{x}}_t^{(j_t)}$
    \STATE Set $b_t = 0$
\ELSE
    \STATE Sample $\bm{x}_t$ uniformly at random from $\{0,1\}^d \setminus \{\bm{x}^{(i)}\}_{i=1}^{t-1}$
    \STATE Set $j_t = \varnothing$
    \STATE Set $b_t = 1$
\ENDIF
\STATE \textbf{Output:} selected point $\bm{x}_t$, selected arm $j_t$, nominated candidates $\{\tilde{\bm{x}}_t^{(m)}\}_{m=1}^{M}$, fallback flag $b_t$
\end{algorithmic}
\end{algorithm}

\begin{algorithm}[!htbp]
\caption{Proposed BOCS with GP-Hedge method}
\label{alg:hybrid}
\begin{algorithmic}[1]
\STATE \textbf{Input:} current dataset $\mathcal{D}_{t-1}$, BOCS surrogate model, Hamming-kernel parameter $\gamma$, LCB parameters $\{\kappa_m\}_{m=1}^{M}$, Hedge parameter $\eta > 0$, cumulative gains $\{g_{t-1}^{(m)}\}_{m=1}^{M}$
\STATE Update the BOCS posterior using $\mathcal{D}_{t-1}$
\STATE Compute candidate $\bm{x}^{\mathrm{BOCS}}_t \approx \arg\min_{\bm{x}\in\{0,1\}^d} \hat{f}_{t-1}(\bm{x})$ using simulated annealing
\STATE Set $b_t = 0$ and $r_t^{(m)} = 0$ for all $m=1,\ldots,M$

\IF{$\bm{x}^{\mathrm{BOCS}}_t \notin \{\bm{x}^{(i)}\}_{i=1}^{t-1}$}
    \STATE Set $\bm{x}_t = \bm{x}^{\mathrm{BOCS}}_t$
\ELSE
    \STATE Compute the GP predictive mean $\mu_{t-1}(\bm{x})$ and standard deviation $\sigma_{t-1}(\bm{x})$ from $\mathcal{D}_{t-1}$ using the Hamming kernel with parameter $\gamma$
    \FOR{$m = 1,\ldots,M$}
        \STATE Define the LCB acquisition function $a_m(\bm{x})=\mu_{t-1}(\bm{x})-\kappa_m\sigma_{t-1}(\bm{x})$
    \ENDFOR
    \STATE Run Algorithm~\ref{alg:gph} and obtain $\bm{x}_t$, $j_t$, $\{\tilde{\bm{x}}_t^{(m)}\}_{m=1}^{M}$, and $b_t$
\ENDIF

\STATE Evaluate $y_t = f(\bm{x}_t)$
\STATE Augment the data $\mathcal{D}_t = \mathcal{D}_{t-1} \cup \{(\bm{x}_t,y_t)\}$

\IF{$\bm{x}^{\mathrm{BOCS}}_t \in \{\bm{x}^{(i)}\}_{i=1}^{t-1}$ and $b_t = 0$}
    \STATE Compute the GP predictive mean $\mu_t(\bm{x})$ from $\mathcal{D}_t$ using the Hamming kernel with parameter $\gamma$
    \FOR{$m = 1,\ldots,M$}
        \STATE Set $r_t^{(m)} = -\mu_t(\tilde{\bm{x}}_t^{(m)})$
    \ENDFOR
\ENDIF

\FOR{$m = 1,\ldots,M$}
    \STATE Update gain $g_t^{(m)} = g_{t-1}^{(m)} + r_t^{(m)}$
\ENDFOR

\STATE \textbf{Output:} updated dataset $\mathcal{D}_t$, updated gains $\{g_t^{(m)}\}_{m=1}^{M}$
\end{algorithmic}
\end{algorithm}

\subsection{Experimental Setup}

As comparison methods, we use GP-Hedge only, which generates search points by GP-Hedge at all iterations without using BOCS, and BOCS with Random, which adds a random point when learning stagnation occurs in BOCS, as in the previous study. The differences among the methods are summarized in Table~\ref{tab:methods}. For all methods, simulated annealing (SA) is used to optimize the acquisition functions.

\begin{table}[!htbp]
\centering
\caption{Comparison methods.}
\label{tab:methods}
\begin{tabular}{ll}
\hline
Method & Stagnation handling \\
\hline
GP-Hedge only & Not applicable \\
BOCS with Random & Add a random point \\
BOCS with GP-Hedge & Add a point by GP-Hedge \\
\hline
\end{tabular}
\end{table}

We use two types of black-box functions: one with the same QUBO structure as the BOCS surrogate model and the other with a more complex higher-order structure. The first benchmark is a fully connected QUBO, defined as
\begin{equation}
f_{\mathrm{QUBO}}(\bm{x})
=
\sum_{i=1}^{d}\alpha_i x_i
+
\sum_{i<j}\alpha_{ij}x_i x_j.
\label{eq:qubo-benchmark}
\end{equation}

The second benchmark is a fully connected Higher-order Unconstrained Binary Optimization (HUBO) problem up to third order, defined as
\begin{equation}
f_{\mathrm{HUBO}}(\bm{x})
=
\sum_{i=1}^{d}\alpha_i x_i
+
\sum_{i<j}\alpha_{ij}x_i x_j
+
\sum_{i<j<k}\alpha_{ijk}x_i x_j x_k.
\label{eq:hubo-benchmark}
\end{equation}

For both benchmarks, the input dimension is set to $d=50$, and the coefficients $\alpha_i$, $\alpha_{ij}$, and $\alpha_{ijk}$ are independently generated from the standard normal distribution $\mathcal{N}(0,1)$. The number of problem instances is 50. As the initial dataset for each benchmark, we generate 50 random 50-dimensional binary vectors in advance. These initial points are shared by all problem instances within the same benchmark. The final iteration $T$ is set to 500 for the fully connected QUBO and 800 for the fully connected HUBO.

In GP-Hedge, we use 10 arms whose hyperparameters in (6) are set to $\kappa_m \in \{1,2,\ldots,10\}$. The Gaussian process uses the Hamming kernel defined in (5). The kernel parameter $\gamma$ is not fixed but updated at each iteration. Specifically, given the observed input set $X_t=\{\bm{x}^{(i)}\}_{i=1}^{t}$, we assume that the observation vector $\bm{y}_t=(y^{(1)},\ldots,y^{(t)})^\top$ follows $\bm{y}_t \sim \mathcal{N}(\bm{0},K_t(\gamma))$, and compute the log marginal likelihood for each candidate value as
\begin{equation}
\log p(\bm{y}_t \mid X_t,\gamma)
=
-\frac{1}{2}\bm{y}_t^\top K_t(\gamma)^{-1}\bm{y}_t
-\frac{1}{2}\log |K_t(\gamma)|
-\frac{t}{2}\log(2\pi).
\label{eq:log-marginal-likelihood}
\end{equation}
Here, $K_t(\gamma)$ is the kernel matrix determined by $\gamma$. In this study, candidate values of $\gamma$ are set on a logarithmic grid from $10^{-3}$ to $10^{0.5}$, and the value that maximizes (8) is selected.

SA is used to optimize the acquisition function corresponding to each arm. At each iteration, we independently run SA 10 times for each arm, and each run consists of 1000 Monte Carlo steps. The best solution in the dataset at that iteration is used as the initial solution. Among the candidates obtained from all runs, the point with the smallest acquisition function value is adopted as the candidate point for that arm. The Hedge parameter used for the probability update in GP-Hedge is fixed to $\eta=1$.

Search performance is evaluated by the Relative gap between the best value obtained up to iteration $t$, $f_{\mathrm{best}}(t):=\min_{(\bm{x},y)\in\mathcal{D}_t} y$, and the reference value $f^\dagger$:
\begin{equation}
\mathrm{Relative\;gap}(t)
=
\frac{f_{\mathrm{best}}(t)-f^\dagger}{|f^\dagger|}.
\label{eq:relative-error}
\end{equation}

For the fully connected QUBO black-box function, the exact optimal value is computed using Gurobi and used as $f^\dagger$ \cite{gurobi}. For the fully connected HUBO, the SA solver in OpenJij is run 1000 times for each instance, and the best value obtained is used as the reference value $f^\dagger$ \cite{openjij}.

As an additional evaluation toward the future use of quantum annealers, we also evaluate the case in which the surrogate model in BOCS is sparsified. In this evaluation, the black-box function is the fully connected QUBO, and the QUBO of the surrogate model used in BOCS is sparsified. Here, sparsity denotes the ratio of off-diagonal elements retained as nonzero elements in the surrogate model, while diagonal elements are always retained. Therefore, a sparsity of 100\% corresponds to a fully connected surrogate model in which all off-diagonal elements are retained, and its structure matches that of the black-box function. In contrast, a sparsity of 0\% corresponds to a surrogate model with only diagonal elements. For each sparsity level, the mask of nonzero elements is randomly generated for each problem instance. However, for the same problem instance and the same sparsity level, the same mask is shared by all comparison methods.

Based on these settings, the next section first presents comparison results for the fully connected QUBO and HUBO benchmarks, and then analyzes the behavior during learning stagnation and the effect of sparse surrogate models.

\section{Results}

\begin{figure*}[!htbp]
    \centering
    \begin{minipage}[t]{0.48\textwidth}
        \centering
        \includegraphics[width=\linewidth]{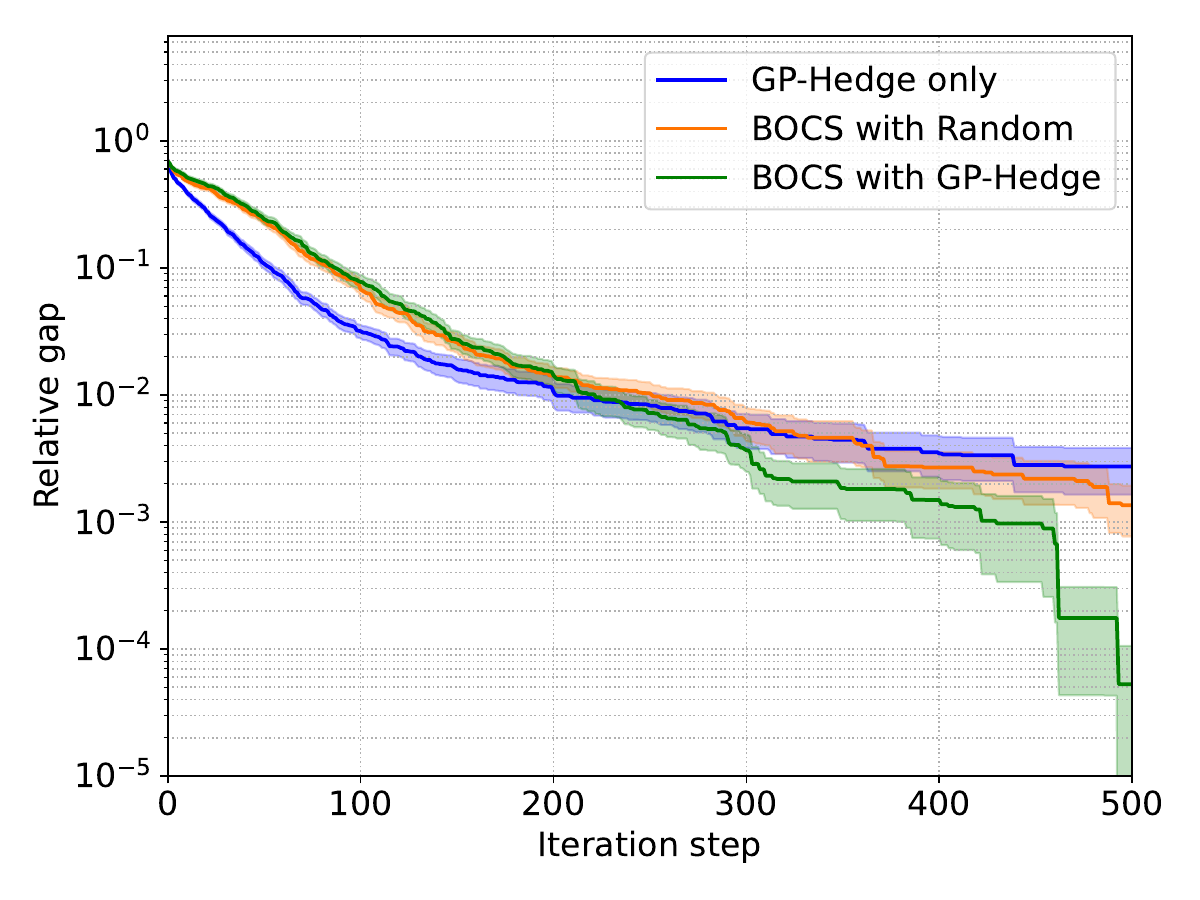}
        \par\smallskip
        {\footnotesize (a) Transition of the Relative gap for each method on the fully connected QUBO benchmark.}
    \end{minipage}
    \hfill
    \begin{minipage}[t]{0.48\textwidth}
        \centering
        \includegraphics[width=\linewidth]{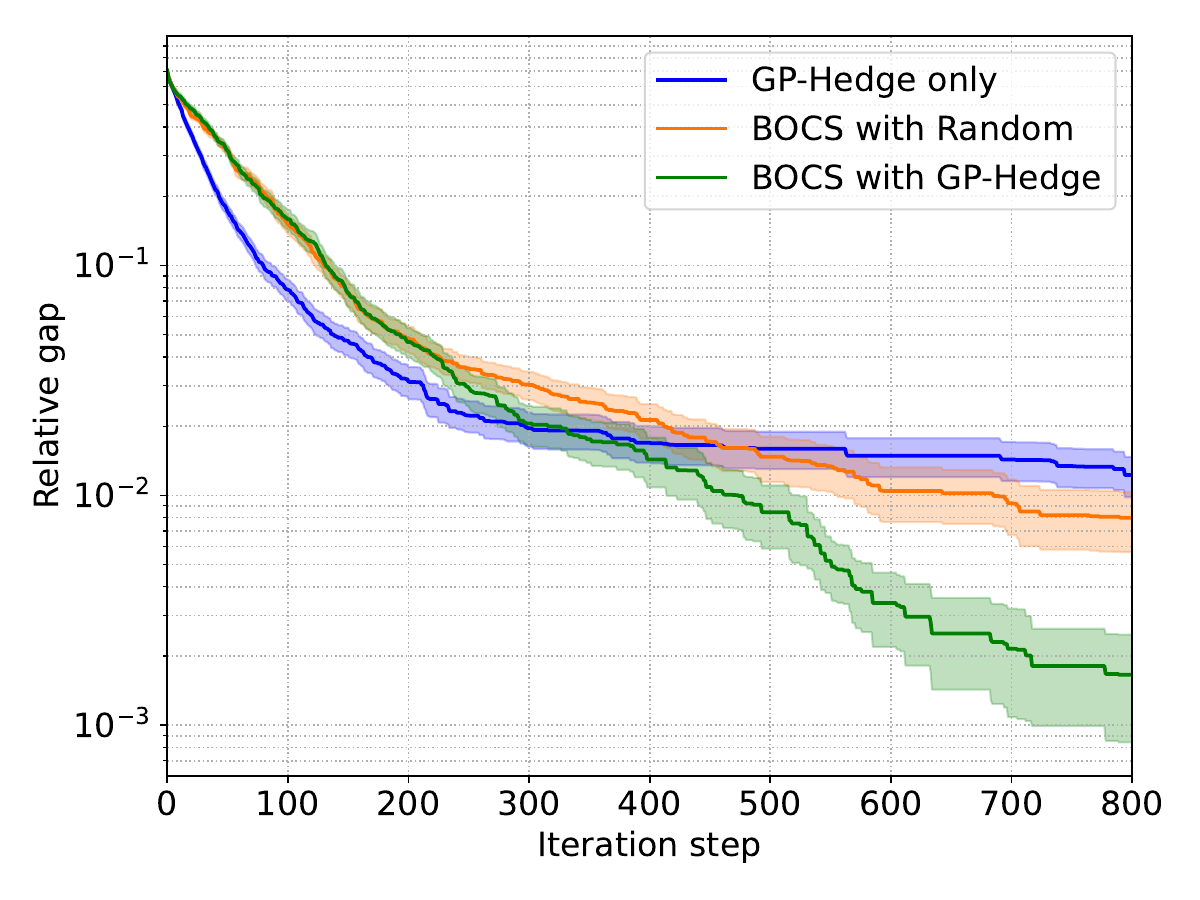}
        \par\smallskip
        {\footnotesize (b) Transition of the Relative gap for each method on the fully connected HUBO benchmark.}
    \end{minipage}
    \caption{Search performance of each method on the fully connected QUBO and HUBO benchmarks. The horizontal axis represents the number of iterations, and the vertical axis represents the Relative gap on a logarithmic scale. The blue, orange, and green lines correspond to GP-Hedge only, BOCS with Random, and the proposed BOCS with GP-Hedge method, respectively. Each curve represents the mean over 50 problem instances, and the shaded region represents the standard error. (a) The proposed method shows behavior similar to BOCS with Random in the first half of the iterations, but reaches the smallest Relative gap in the middle and later stages. (b) Even when the black-box function is more complex than the BOCS surrogate model, the proposed method reaches the smallest final Relative gap.}
    \label{fig:blackbox}
\end{figure*}

We first present the results for the fully connected QUBO black-box function. As shown in Fig.~\ref{fig:blackbox}(a), the proposed BOCS with GP-Hedge method exhibits almost the same behavior as BOCS with Random in the first half of the search. This is natural because both methods run BOCS until learning stagnation occurs. In contrast, in the middle and later stages, BOCS with GP-Hedge reaches a smaller Relative gap. GP-Hedge only decreases the Relative gap more rapidly than the other methods in the early stage. This behavior is considered to occur because the number of observations is small in the early stage, and the predictive uncertainty of the Gaussian process is large for many unevaluated points. As a result, the uncertainty term in (6) has a relatively strong effect, and search points are likely to be selected from a wide candidate region. Consequently, GP-Hedge only reaches the neighborhood of a promising region at an early stage, and the Relative gap temporarily decreases rapidly. However, it does not sufficiently exploit the promising region afterward, and its final performance is inferior to that of the proposed method. These results indicate that the proposed method, which mainly uses BOCS when the amount of data is small and activates GP-Hedge only when learning stagnation occurs, works effectively over the entire search process.

Next, Fig.~\ref{fig:blackbox}(b) shows the results for the fully connected HUBO black-box function. As in the fully connected QUBO case, the differences among the three methods are relatively small in the first half of the iterations, whereas the superiority of the proposed method becomes clear in the middle and later stages. This result indicates that the proposed method can improve learning stagnation and perform efficient search even when the black-box function is more complex than the BOCS surrogate model.

Table~\ref{tab:mean-reduction} shows the improvement of the proposed method in Fig.~\ref{fig:blackbox}. Let $\bar{f}_{m}(T)$ be the instance-averaged Relative gap of method $m$ at the final iteration $T$, and let $\bar{f}_{\mathrm{ours}}(T)$ be the result of the proposed BOCS with GP-Hedge method. The final performance improvement over method $m$ is defined as
\begin{equation}
\mathrm{Value\ improvement}_{m} \; (\%)
=
\frac{\bar{f}_{m}(T)-\bar{f}_{\mathrm{ours}}(T)}{\bar{f}_{m}(T)}
\times 100 .
\label{eq:mean-error-reduction}
\end{equation}

For the fully connected QUBO, the proposed method improves the final Relative gap by $98.07\%$ over GP-Hedge only and by $96.08\%$ over BOCS with Random. For the fully connected HUBO, the corresponding improvements are $86.46\%$ and $79.25\%$, respectively.

We next quantitatively evaluate how much the proposed method reduces the number of evaluations required to reach the final performance of each comparison method. For instance $i$, let $f_m^{(i)}(T)$ be the Relative gap of comparison method $m$ at the final iteration, and let $T_m^{r,(i)}$ be the first iteration at which the proposed method reaches a Relative gap no larger than $f_m^{(i)}(T)$. Because the proposed method may not reach this value by the final iteration, we define the iteration improvement as
\begin{equation}
\begin{aligned}
&\mathrm{Iteration\ improvement}_{m}^{(i)} \; (\%) \\
&=
\begin{cases}
\left(1-\dfrac{T_{m}^{r,(i)}}{T}\right)\times 100, 
& \begin{array}{l}
\text{if } T_{m}^{r,(i)}
\text{exists},
\end{array}\\[8pt]
0, & \text{otherwise}.
\end{cases}
\end{aligned}
\label{eq:iteration-improvement}
\end{equation}

When the proposed method does not reach the target value, the improvement is set to $0$, meaning that no reduction in the number of evaluations is confirmed. Table~\ref{tab:attainment} shows the instance averages and standard errors. For the fully connected QUBO, the proposed method reduces the number of evaluations by $53.1\%$ compared with GP-Hedge only and by $49.1\%$ compared with BOCS with Random. For the fully connected HUBO, the corresponding reductions are $55.1\%$ and $53.4\%$, respectively. The success counts for reaching $f_m^{(i)}(T)$ are $50/50$ for QUBO and $48/50$ for HUBO, indicating that the proposed method reaches the final performance of the comparison methods with fewer evaluations in most instances.

\begin{table*}[!htbp]
\centering
\caption{Final performance improvement of the proposed method at the final iteration $T$.}
\label{tab:mean-reduction}
\begin{tabular}{ll||cc}
\hline
Black-box function & Method & $\bar{f}_{m}(T)$ & Value improvement (\%) \\
\hline
\multirow{3}{*}{QUBO}
& GP-Hedge only & $2.737\times 10^{-3}$ & $98.07$ \\
& BOCS with Random & $1.352\times 10^{-3}$ & $96.08$ \\
& BOCS with GP-Hedge & $5.292\times 10^{-5}$ & -- \\
\hline
\multirow{3}{*}{HUBO}
& GP-Hedge only & $1.226\times 10^{-2}$ & $86.46$ \\
& BOCS with Random & $8.000\times 10^{-3}$ & $79.25$ \\
& BOCS with GP-Hedge & $1.660\times 10^{-3}$ & -- \\
\hline
\end{tabular}
\end{table*}

\begin{table*}[!htbp]
\centering
\caption{Reduction in the number of evaluations achieved by the proposed method.}
\label{tab:attainment}
\small
\setlength{\tabcolsep}{4pt}
\begin{tabular}{l||cc|cc}
\hline
\multirow{2}{*}{Method} & \multicolumn{2}{c}{QUBO} & \multicolumn{2}{c}{HUBO} \\
\cline{2-5}
& \begin{tabular}[c]{@{}c@{}}Iteration\\ improvement (\%)\end{tabular} & \begin{tabular}[c]{@{}c@{}}Success\\ count\end{tabular} & \begin{tabular}[c]{@{}c@{}}Iteration\\ improvement (\%)\end{tabular} & \begin{tabular}[c]{@{}c@{}}Success\\ count\end{tabular} \\
\hline
GP-Hedge only & $53.1 \pm 2.4$ & $50/50$ & $55.1 \pm 3.3$ & $48/50$ \\
BOCS with Random & $49.1 \pm 2.9$ & $50/50$ & $53.4 \pm 3.2$ & $48/50$ \\
\hline
\end{tabular}
\end{table*}

\subsection{Behavior Around Learning Stagnation in Fully Connected QUBO}

\begin{figure*}[!htbp]
    \centering
        \begin{minipage}[t]{0.48\textwidth}
        \centering
        \includegraphics[width=\linewidth]{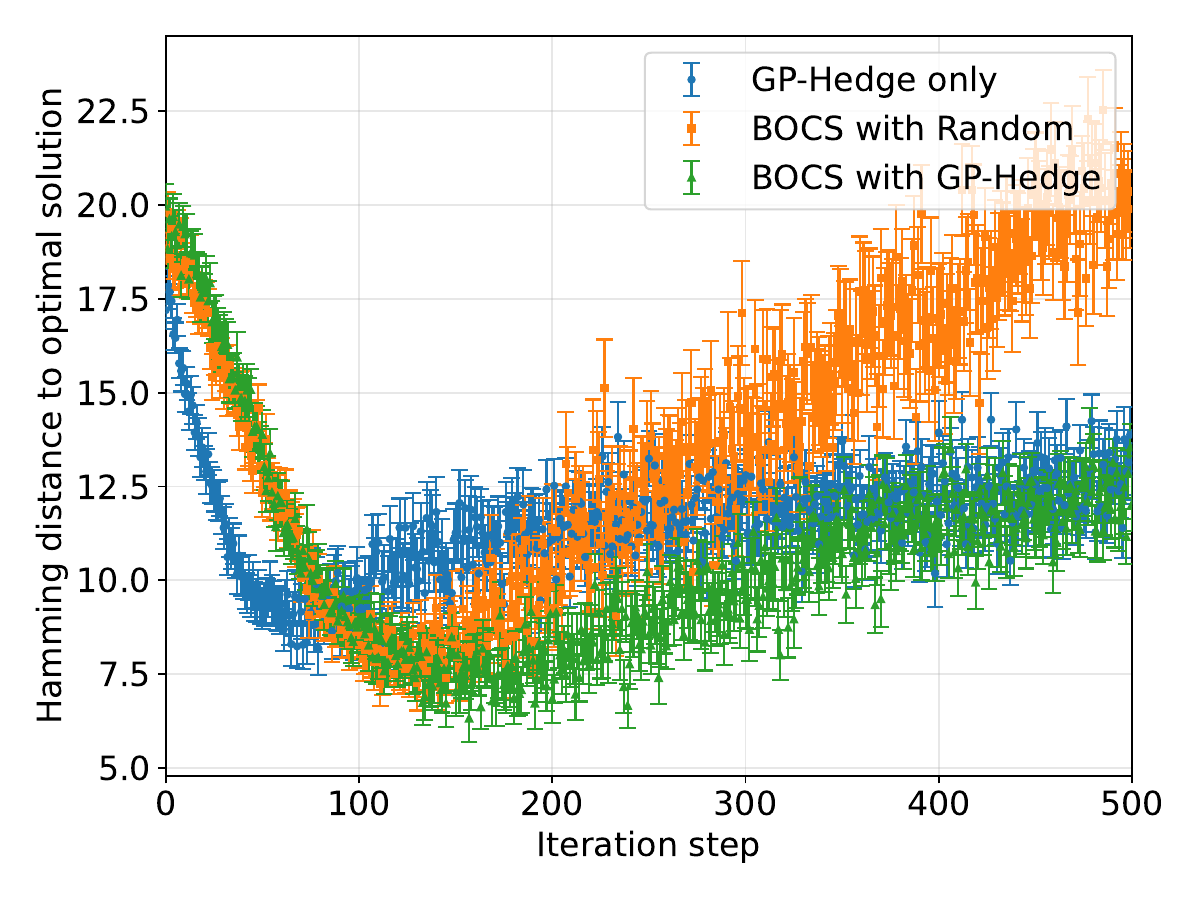}
        \par\smallskip
        {\footnotesize (a) Transition of the Hamming distance between the point proposed at each iteration and the optimal solution.}
    \end{minipage}
    \hfill
        \begin{minipage}[t]{0.48\textwidth}
        \centering
        \includegraphics[width=\linewidth]{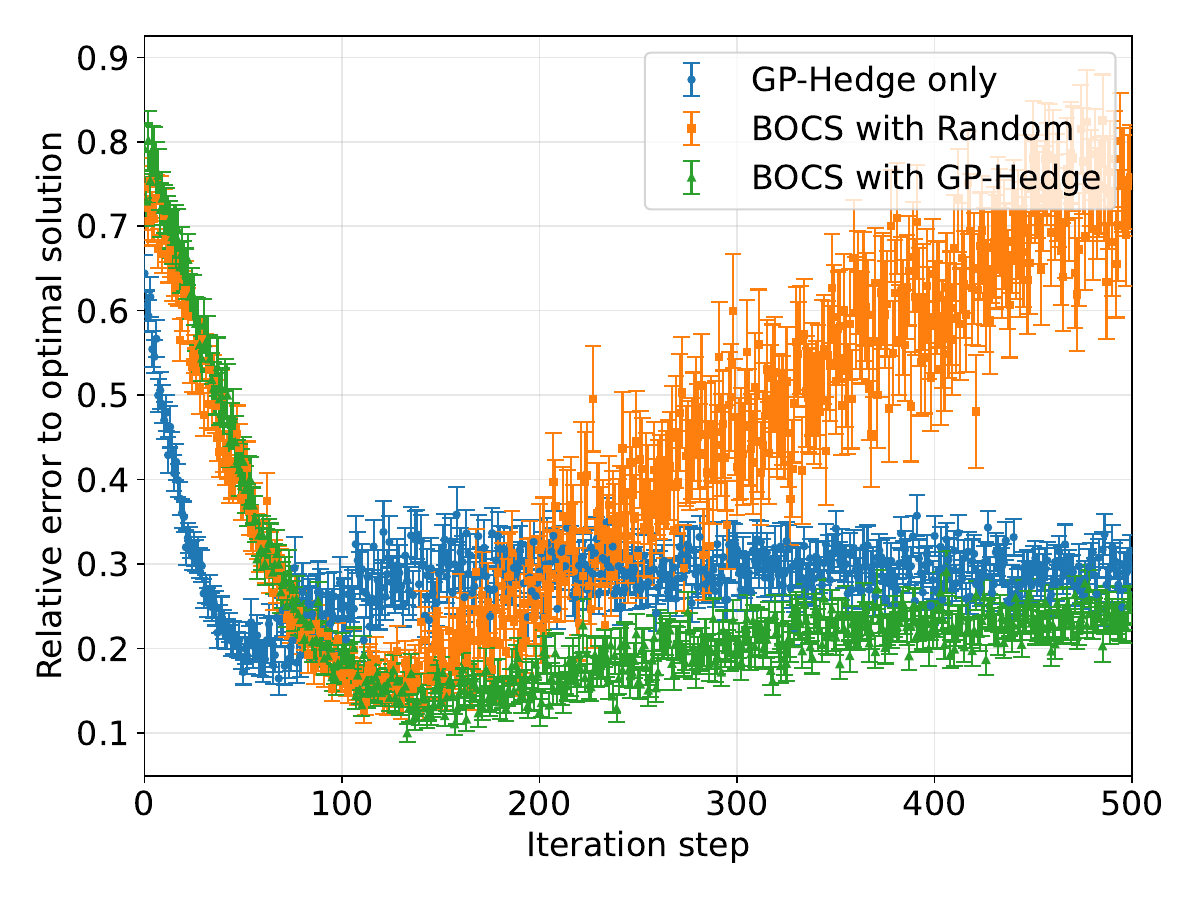}
        \par\smallskip
        {\footnotesize (b) Transition of the relative error between the objective value of the point proposed at each iteration and the optimal value.}
    \end{minipage}
        \caption{Analysis of proposed points on the fully connected QUBO benchmark. The horizontal axis represents the number of iterations, and the vertical axis represents the Hamming distance to the optimal solution or the relative error of the objective value. The blue, orange, and green lines correspond to GP-Hedge only, BOCS with Random, and the proposed BOCS with GP-Hedge method, respectively. Each curve represents the mean over 50 problem instances, and the error bars represent the standard error. The proposed method provides points with a Hamming distance similar to that of GP-Hedge only in the later stage, while achieving smaller relative errors.}
    \label{fig:stagnation}
\end{figure*}

To examine how the proposed method avoids learning stagnation, Fig.~\ref{fig:stagnation} shows the transition of the Hamming distance between the point proposed by the acquisition function and the optimal solution at each iteration, as well as the relative error of the objective value of the proposed point with respect to the optimal value, for the fully connected QUBO benchmark. As shown in Fig.~\ref{fig:stagnation}(a), GP-Hedge only rapidly decreases the Hamming distance to the optimal solution in the early stage, but the decrease slows afterward. BOCS with Random also decreases the distance until the middle stage, but the distance increases again in the later stage because random-point addition occurs more frequently. In contrast, BOCS with GP-Hedge achieves the smallest Hamming distance in the middle stage and eventually reaches a distance comparable to that of GP-Hedge only. Taken together, these results suggest that the proposed method accumulates observations near promising regions by using BOCS before activating GP-Hedge, whereas GP-Hedge only does not sufficiently narrow down the search region in the early stage. This difference leads to the performance gap in the later stage. In other words, the quality of the observations already obtained when GP-Hedge is activated is important.

\subsection{Sparse Surrogate Model}

\begin{figure*}[!htbp]
    \centering
    \begin{minipage}[t]{0.48\textwidth}
        \centering
        \includegraphics[width=\linewidth]{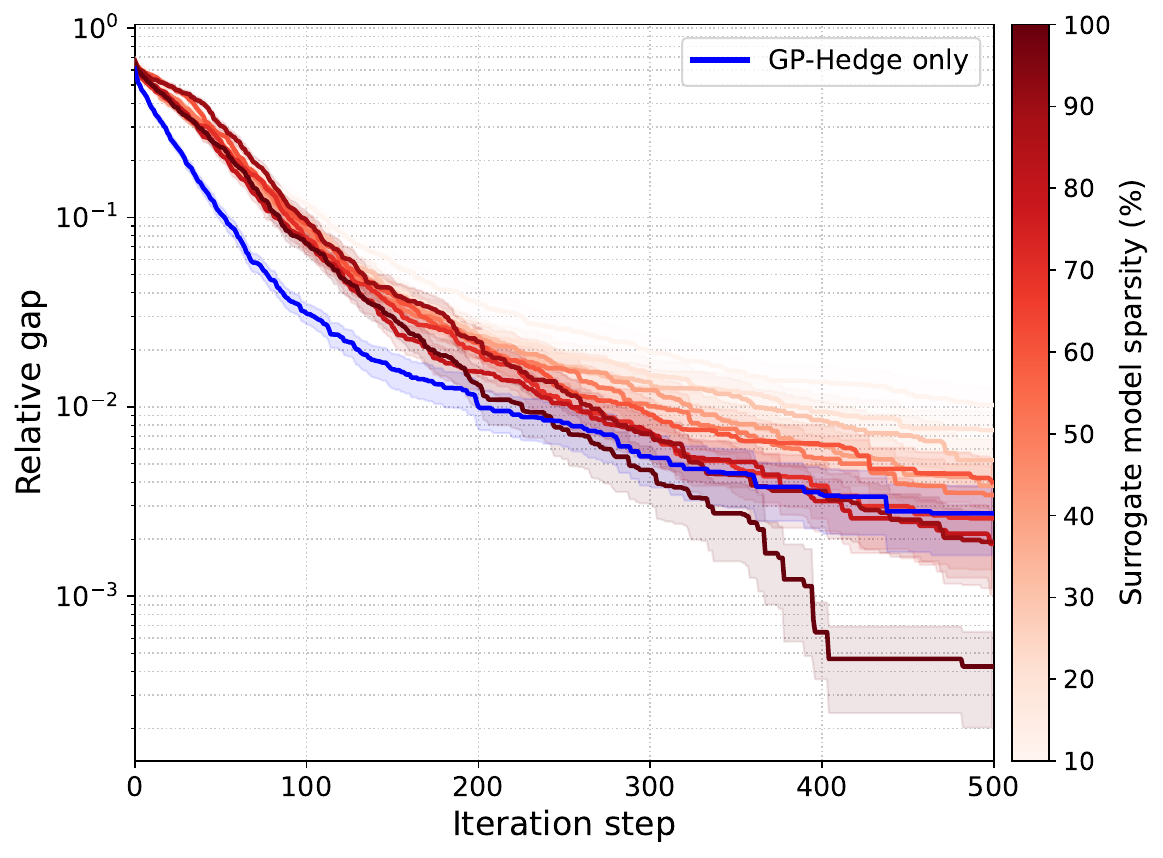}
        \par\smallskip
        {\footnotesize (a) Transition of the Relative gap for different sparsity levels.}
    \end{minipage}
    \hfill
    \begin{minipage}[t]{0.48\textwidth}
        \centering
        \includegraphics[width=\linewidth]{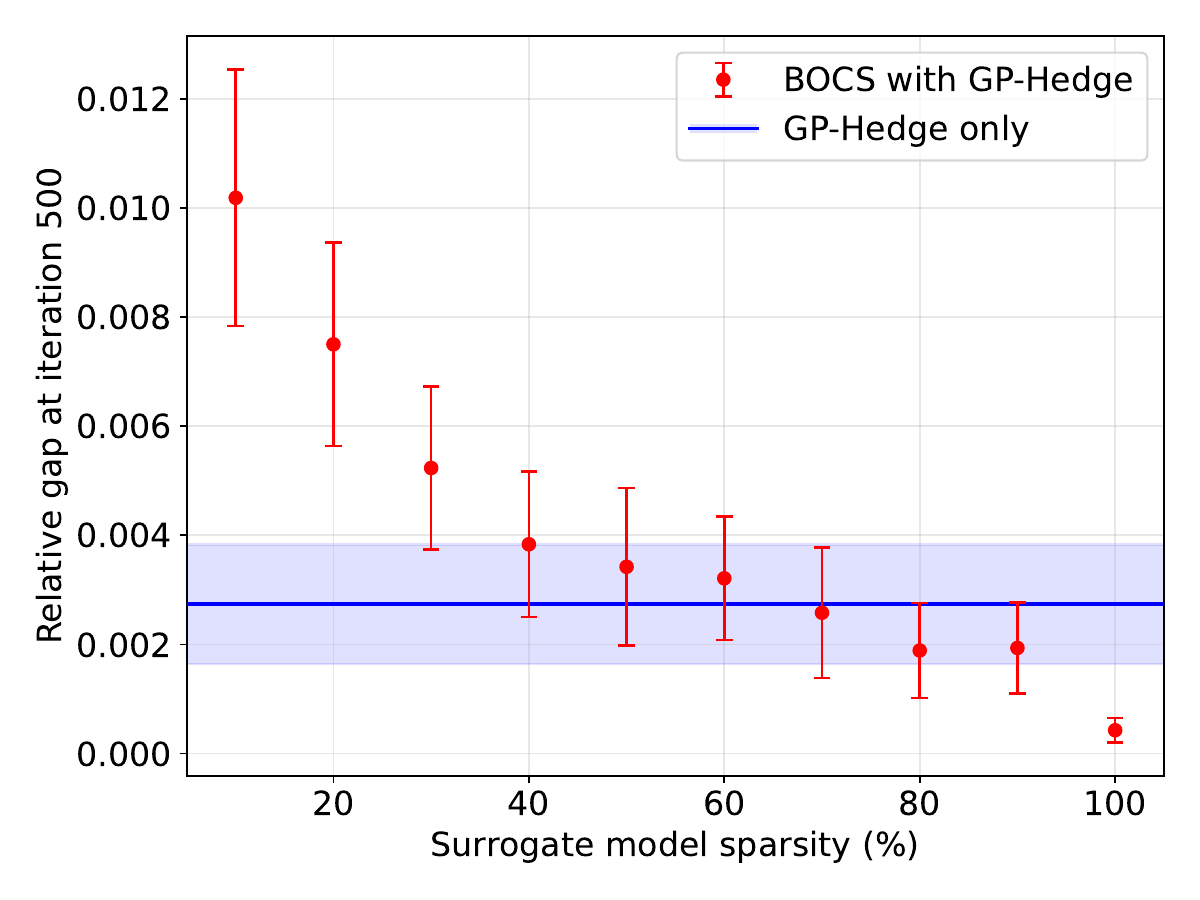}
        \par\smallskip
        {\footnotesize (b) Relative gap at 500 iterations for each sparsity level.}
    \end{minipage}
    \caption{Effect of sparsifying the BOCS surrogate model on the fully connected QUBO benchmark. (a) The horizontal axis represents the number of iterations, and the vertical axis represents the Relative gap. Sparsity denotes the ratio of nonzero off-diagonal elements in the surrogate model, while diagonal elements are always retained. A sparsity of 100\% corresponds to the fully connected surrogate model, and a sparsity of 0\% corresponds to a surrogate model with only diagonal elements. The blue line shows the result of GP-Hedge only, and the red curves show the results of BOCS with GP-Hedge. Each curve represents the mean over 50 problem instances, and the shaded region represents the standard error. (b) The horizontal axis represents sparsity, and the vertical axis represents the Relative gap at 500 iterations. The red points and error bars represent the mean and standard error of BOCS with GP-Hedge, whereas the blue line and band represent the mean and standard error of GP-Hedge only.}
    \label{fig:sparse}
\end{figure*}

Finally, we present an additional evaluation of sparse surrogate models toward the future use of quantum annealers. As shown in Fig.~\ref{fig:sparse}(a), the convergence speed and final accuracy of the proposed method deteriorate as the sparsity decreases, that is, as the surrogate model becomes sparser. Even at a sparsity of 90\%, the performance of the proposed method remains comparable to that of GP-Hedge only. This result indicates that the surrogate model must retain connections close to those of the fully connected model to fully exploit the performance of the proposed method. Fig.~\ref{fig:sparse}(b) summarizes the Relative gap at the final iteration as a function of sparsity, and this trend is more clearly observed.

This result indicates that sparsifying the surrogate model reduces its representational capacity, and that this loss is difficult to compensate for by GP-Hedge. As shown in Fig.~\ref{fig:blackbox}(a), for the proposed method to work effectively, BOCS must accumulate sufficiently promising observations before GP-Hedge is activated. However, when the surrogate model is sparsified, its limited representational capacity prevents BOCS from finding sufficiently promising points before GP-Hedge is activated. As a result, the proposed method becomes less effective.

\section{Discussion and Conclusion}

To examine whether the effectiveness of the proposed method comes simply from exploring the neighborhood of promising points or from adaptively selecting evaluation points within that neighborhood, we compare it with BOCS with SpinFlip. In BOCS with SpinFlip, when learning stagnation occurs in BOCS, an unevaluated point obtained by randomly applying one spin flip to the current best point is added. If all points reachable by one spin flip have already been evaluated, the search range is expanded to two-spin-flip and then three-spin-flip neighborhoods until an unevaluated point is selected.

\begin{figure*}[!htbp]
    \centering
    \begin{minipage}[t]{0.48\textwidth}
        \centering
        \includegraphics[width=\linewidth]{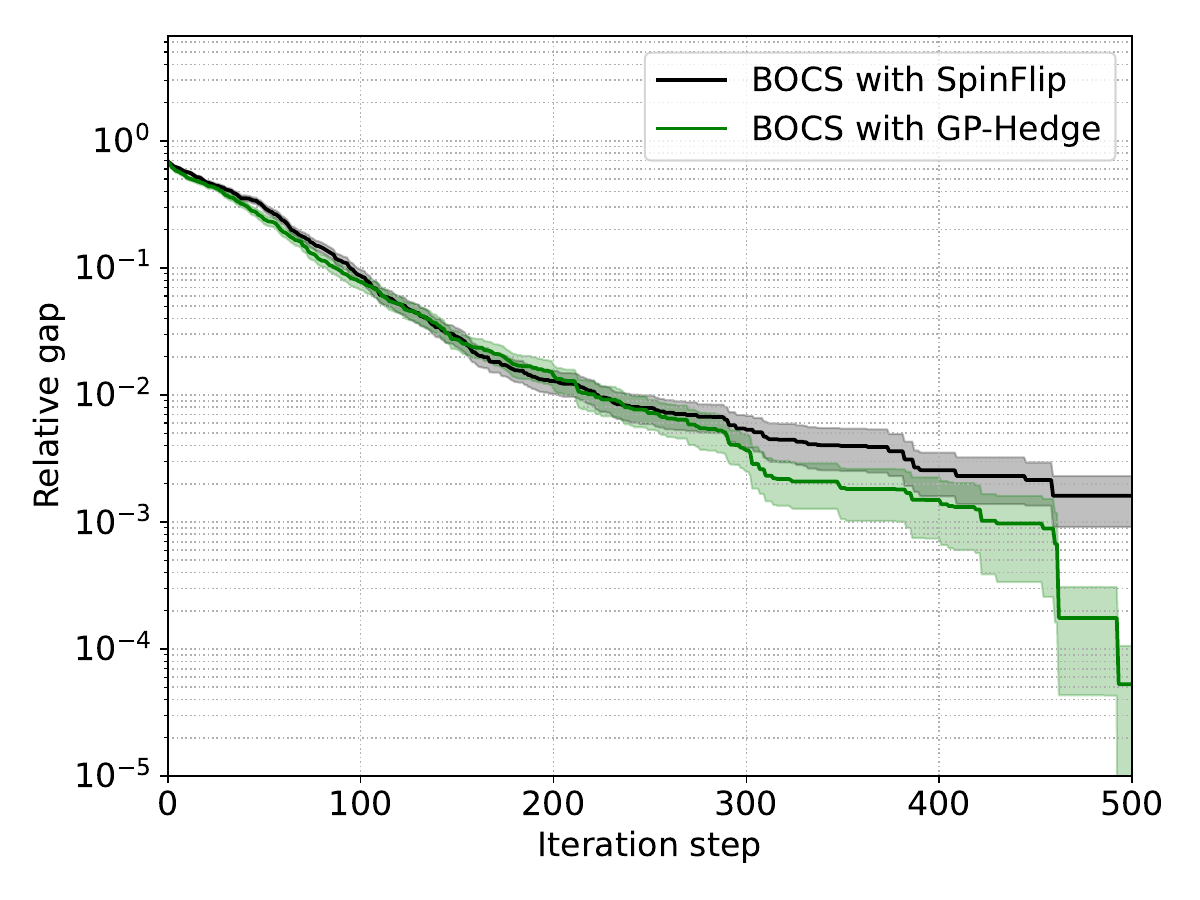}
        \par\smallskip
        {\footnotesize (a) Comparison of Relative gap with BOCS with SpinFlip.}
    \end{minipage}
    \hfill
    \begin{minipage}[t]{0.48\textwidth}
        \centering
        \includegraphics[width=\linewidth]{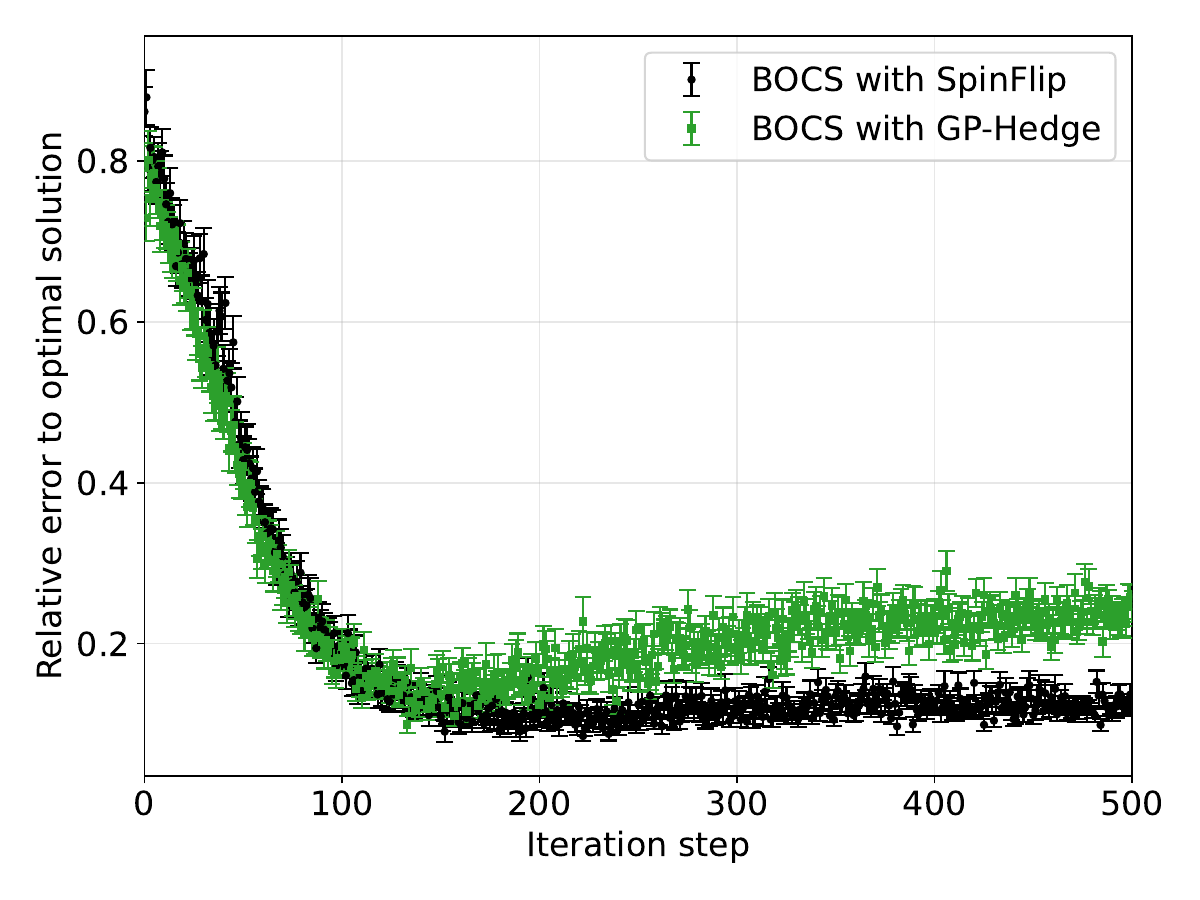}
        \par\smallskip
        {\footnotesize (b) Comparison of the Relative gap of proposed points.}
    \end{minipage}

    \vspace{0.8em}

    \begin{minipage}[t]{0.48\textwidth}
        \centering
        \includegraphics[width=\linewidth]{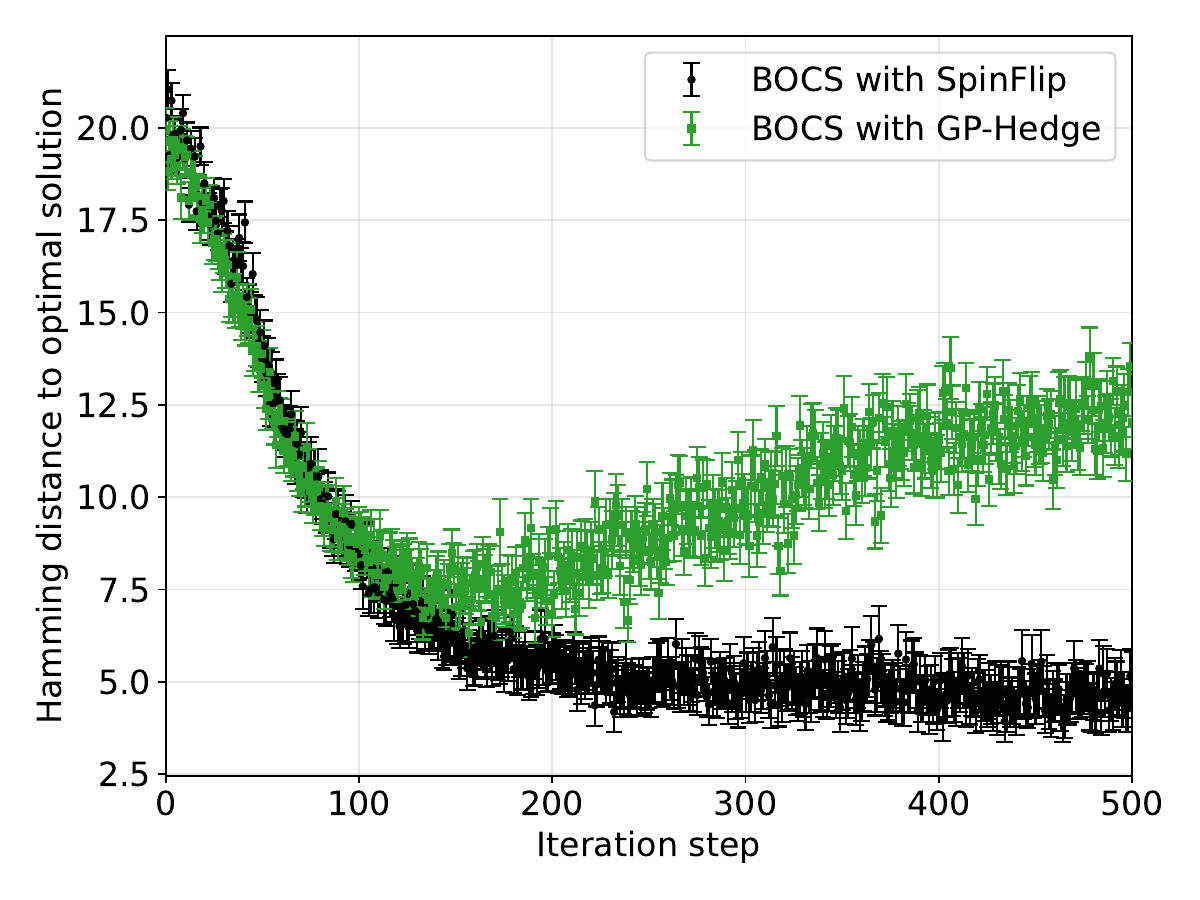}
        \par\smallskip
        {\footnotesize (c) Comparison of the Hamming distance between proposed points.}
    \end{minipage}

    \caption{Comparison with BOCS with SpinFlip on the fully connected QUBO benchmark. (a) The horizontal axis represents the number of iterations, and the vertical axis represents the Relative gap. (b) The horizontal axis represents the number of iterations, and the vertical axis represents the Relative gap between the objective value of the point proposed at each iteration and the optimal value. (c) The horizontal axis represents the number of iterations, and the vertical axis represents the Hamming distance between the point proposed at each iteration and the point proposed at the previous iteration. Each curve or point represents the mean over 50 problem instances, and the shaded regions or error bars represent the standard error. These results indicate that the effectiveness of the proposed method does not come simply from adding low-value points near the current best point.}
    \label{fig:discussion-analysis}
\end{figure*}

Fig.~\ref{fig:discussion-analysis}(a) shows the transition of the Relative gap for BOCS with SpinFlip and the proposed BOCS with GP-Hedge method. The two methods show similar behavior until the middle stage, but the proposed method reaches a smaller Relative gap in the later stage. On the other hand, Fig.~\ref{fig:discussion-analysis}(b) and Fig.~\ref{fig:discussion-analysis}(c) show that the points obtained by spin flips have lower objective values and smaller Hamming distances than those proposed by the proposed method. Therefore, simply adding low-value points near the current best point during learning stagnation is not sufficient to improve the overall search performance.

In this study, we proposed BOCS with GP-Hedge to improve learning stagnation in BOCS for discrete-variable black-box optimization. The proposed method performs the standard BOCS search when BOCS can propose an unevaluated point, and generates an alternative point using a Gaussian process and GP-Hedge only when BOCS proposes an already evaluated point. This design aims to preserve the efficiency of BOCS under limited data while exploiting the observed data to generate unevaluated and promising search points during learning stagnation.

Numerical experiments using fully connected QUBO and fully connected HUBO benchmarks showed that the proposed method reached smaller Relative gaps than GP-Hedge only and the existing random-point addition method. The proposed method also reached the final performance of the comparison methods with fewer evaluations, confirming the effectiveness of replacing random-point addition during learning stagnation with GP-Hedge-based point generation. In particular, the proposed method improved performance even for HUBO, whose black-box function is more complex than the quadratic surrogate model of BOCS. This result suggests that the proposed method can work effectively even when the representational limitation of the surrogate model becomes apparent.

Additional analysis confirmed that the effectiveness of the proposed method does not come simply from adding low-value points near the current best point. The comparison with BOCS with SpinFlip suggests that mechanically adding neighborhood points provides only limited improvement, and that selecting additional points based on the predictive mean and uncertainty of the Gaussian process is important.

On the other hand, experiments with sparse BOCS surrogate models for future use with quantum annealers showed that even slight sparsification prevents the proposed method from improving performance. This result indicates that, to fully exploit the proposed method, the surrogate model must have sufficient representational capacity for BOCS to obtain promising observations before learning stagnation occurs.

In this study, the candidate acquisition functions in GP-Hedge were limited to LCB, and multiple arms were constructed by setting the hyperparameters to $\kappa_m=1,\ldots,10$. Compared with using a single acquisition function, this setting makes tuning easier because the hyperparameter does not need to be fixed to a single value and can instead be specified as a range. However, developing a method for automatically determining an appropriate range remains future work. Another future direction is to examine how performance changes when other acquisition functions, such as Probability of Improvement (PI) and Expected Improvement (EI), or acquisition functions that do not rely on Gaussian processes, are used as candidates.

\newpage

\bibliographystyle{IEEEtran}
\bibliography{mybibliography}

@inproceedings{baptista2018bocs,
  title     = {Bayesian Optimization of Combinatorial Structures},
  author    = {Baptista, Ricardo and Poloczek, Matthias},
  booktitle = {Proceedings of the 35th International Conference on Machine Learning},
  series    = {Proceedings of Machine Learning Research},
  volume    = {80},
  pages     = {462--471},
  year      = {2018},
  editor    = {Dy, Jennifer and Krause, Andreas},
  publisher = {PMLR},
  url       = {https://proceedings.mlr.press/v80/baptista18a.html}
}

@book{rasmussen2006gpml,
  title     = {Gaussian Processes for Machine Learning},
  author    = {Rasmussen, Carl Edward and Williams, Christopher K. I.},
  publisher = {The MIT Press},
  address   = {Cambridge, MA},
  year      = {2006},
  isbn      = {9780262182539}
}

@inproceedings{hoffman2011portfolio,
  title     = {Portfolio Allocation for Bayesian Optimization},
  author    = {Hoffman, Matthew W. and Brochu, Eric and de Freitas, Nando},
  booktitle = {Proceedings of the 27th Conference on Uncertainty in Artificial Intelligence},
  pages     = {327--336},
  year      = {2011},
  publisher = {AUAI Press},
  url       = {https://arxiv.org/abs/1009.5419}
}

@article{morita2023random,
  title   = {Random Postprocessing for Combinatorial Bayesian Optimization},
  author  = {Morita, Keisuke and Nishikawa, Yoshihiko and Ohzeki, Masayuki},
  journal = {Journal of the Physical Society of Japan},
  volume  = {92},
  number  = {12},
  pages   = {123801},
  year    = {2023},
  doi     = {10.7566/JPSJ.92.123801},
  url     = {https://doi.org/10.7566/JPSJ.92.123801}
}

@article{kadowaki1998qa,
  title   = {Quantum Annealing in the Transverse Ising Model},
  author  = {Kadowaki, Tadashi and Nishimori, Hidetoshi},
  journal = {Physical Review E},
  volume  = {58},
  number  = {5},
  pages   = {5355--5363},
  year    = {1998},
  doi     = {10.1103/PhysRevE.58.5355},
  url     = {https://doi.org/10.1103/PhysRevE.58.5355}
}

@article{johnson2011qa,
  title   = {Quantum Annealing with Manufactured Spins},
  author  = {Johnson, M. W. and Amin, M. H. S. and Gildert, Suzanne and Lanting, Trevor and Hamze, Firas and Dickson, Neil and Harris, Richard and Berkley, Andrew J. and Johansson, Jan and Bunyk, Paul and Chapple, Elizabeth M. and Enderud, Jeremy and Hilton, Jason P. and Karimi, Kelly and Ladizinsky, Evgeny and Ladizinsky, Nam and Oh, Tomas and Perminov, Igor and Rich, Colin and Thom, M. C. and Tolkacheva, Elena and Truncik, C. J. S. and Uchaikin, Sergey},
  journal = {Nature},
  volume  = {473},
  number  = {7346},
  pages   = {194--198},
  year    = {2011},
  doi     = {10.1038/nature10012},
  url     = {https://doi.org/10.1038/nature10012}
}

@article{koshikawa2021dwavebbo,
  title   = {Benchmark Test of Black-box Optimization Using {D-Wave} Quantum Annealer},
  author  = {Koshikawa, Ami S. and Ohzeki, Masayuki and Kadowaki, Tadashi and Tanaka, Kazuyuki},
  journal = {Journal of the Physical Society of Japan},
  volume  = {90},
  number  = {6},
  pages   = {064001},
  year    = {2021},
  doi     = {10.7566/JPSJ.90.064001},
  url     = {https://doi.org/10.7566/JPSJ.90.064001}
}

@article{doi2023chemical,
  title   = {Exploration of New Chemical Materials Using Black-box Optimization with the {D-Wave} Quantum Annealer},
  author  = {Doi, Mikiya and Nakao, Yoshihiro and Tanaka, Takuro and Sako, Masami and Ohzeki, Masayuki},
  journal = {Frontiers in Computer Science},
  volume  = {5},
  pages   = {1286226},
  year    = {2023},
  doi     = {10.3389/fcomp.2023.1286226},
  url     = {https://doi.org/10.3389/fcomp.2023.1286226}
}

@article{otsuka2025filtering,
  title   = {Filtering out Mislabeled Training Instances Using Black-box Optimization and Quantum Annealing},
  author  = {Otsuka, Makoto and Kodama, Kento and Morita, Keisuke and Ohzeki, Masayuki},
  journal = {Scientific Reports},
  volume  = {15},
  pages   = {37892},
  year    = {2025},
  doi     = {10.1038/s41598-025-21686-z},
  url     = {https://doi.org/10.1038/s41598-025-21686-z}
}

@article{choi2008minor,
  title   = {Minor-Embedding in Adiabatic Quantum Computation: {I}. The Parameter Setting Problem},
  author  = {Choi, Vicky},
  journal = {Quantum Information Processing},
  volume  = {7},
  number  = {5},
  pages   = {193--209},
  year    = {2008},
  doi     = {10.1007/s11128-008-0082-9},
  url     = {https://doi.org/10.1007/s11128-008-0082-9}
}

@article{tasseff2024emerging,
  title   = {On the Emerging Potential of Quantum Annealing Hardware for Combinatorial Optimization},
  author  = {Tasseff, Byron and Albash, Tameem and Morrell, Zachary and Vuffray, Marc and Lokhov, Andrey Y. and Misra, Sidhant and Coffrin, Carleton},
  journal = {Journal of Heuristics},
  volume  = {30},
  number  = {5--6},
  pages   = {325--358},
  year    = {2024},
  doi     = {10.1007/s10732-024-09530-5},
  url     = {https://doi.org/10.1007/s10732-024-09530-5}
}

@article{kirkpatrick1983optimization,
  author  = {Kirkpatrick, Scott and Gelatt, C. Daniel and Vecchi, Mario P.},
  title   = {Optimization by Simulated Annealing},
  journal = {Science},
  year    = {1983},
  volume  = {220},
  number  = {4598},
  pages   = {671--680},
  doi     = {10.1126/science.220.4598.671}
}

@article{kitai2020designing,
  title = {Designing metamaterials with quantum annealing and factorization machines},
  author = {Kitai, Koki and Guo, Jiang and Ju, Shenghong and Tanaka, Shu and Tsuda, Koji and Shiomi, Junichiro and Tamura, Ryo},
  journal = {Physical Review Research},
  volume = {2},
  number = {1},
  pages = {013319},
  year = {2020},
  doi = {10.1103/PhysRevResearch.2.013319}
}

@article{tucs2023quantum,
  title = {Quantum Annealing Designs Nonhemolytic Antimicrobial Peptides in a Discrete Latent Space},
  author = {Tu{\v{c}}s, Andrejs and Berenger, Francois and Yumoto, Akiko and Tamura, Ryo and Uzawa, Takanori and Tsuda, Koji},
  journal = {ACS Medicinal Chemistry Letters},
  volume = {14},
  pages = {577--582},
  year = {2023},
  doi = {10.1021/acsmedchemlett.3c00094}
}

@article{shahriari2016taking,
  author  = {Shahriari, Bobak and Swersky, Kevin and Wang, Ziyu and Adams, Ryan P. and de Freitas, Nando},
  title   = {Taking the Human Out of the Loop: A Review of Bayesian Optimization},
  journal = {Proceedings of the IEEE},
  year    = {2016},
  volume  = {104},
  number  = {1},
  pages   = {148--175},
  doi     = {10.1109/JPROC.2015.2494218}
}

@article{jones1998efficient,
  author  = {Jones, Donald R. and Schonlau, Matthias and Welch, William J.},
  title   = {Efficient Global Optimization of Expensive Black-Box Functions},
  journal = {Journal of Global Optimization},
  year    = {1998},
  volume  = {13},
  number  = {4},
  pages   = {455--492},
  doi     = {10.1023/A:1008306431147}
}

@inproceedings{srinivas2010gpucb,
  title     = {Gaussian Process Optimization in the Bandit Setting: No Regret and Experimental Design},
  author    = {Srinivas, Niranjan and Krause, Andreas and Kakade, Sham M. and Seeger, Matthias},
  booktitle = {Proceedings of the 27th International Conference on Machine Learning},
  pages     = {1015--1022},
  year      = {2010}
}

@misc{gurobi,
  author = {Gurobi Optimization, LLC},
  title  = {Gurobi Optimizer Reference Manual},
  year   = {2026},
  url    = {https://www.gurobi.com}
}

@misc{openjij,
  author = {Kohji Nishimura and Yoshiki Sakamoto and Taro Shimizu and Kohei Suzuki and Yu Yamashiro},
  title  = {OpenJij},
  year   = {2025},
  month  = oct,
  note   = {Version 0.11.6},
  url    = {https://github.com/Jij-Inc/OpenJij}
}

\newpage

\EOD

\end{document}